\documentclass[11pt]{article}
\usepackage{amsmath,paralist,amsthm,comment}
\usepackage{cite}

\DeclareMathOperator{\tr}{tr}
\DeclareMathOperator{\Tr}{Tr}

\DeclareMathOperator{\Irr}{Irr}
\DeclareMathOperator{\supp}{supp}

\DeclareMathOperator{\swt}{swt}
\DeclareMathOperator{\wt}{wt}

\newcommand{\C}{\mathbf{C}}
\newcommand{\F}{\mathbf{F}}
\newcommand{\nix}[1]{}
\newcommand{\sdual}{{\perp_s}}
\newcommand{\adual}{{\perp_a}}
\newcommand{\ol}{\overline}
\newcommand{\ket}[1]{|#1\rangle}

\newtheorem{theorem}{Theorem}

\newtheorem{lemma}[theorem]{Lemma}
\newtheorem{fact}{Fact}

\begin{document}
\begin{center}
\textbf{\Large Clifford Code Constructions of\\[0.5ex]
Operator Quantum Error-Correcting Codes}\\[1ex]
Andreas Klappenecker and Pradeep Kiran Sarvepalli\\
Department of Computer Science, Texas A\&M University
\end{center}
\begin{abstract}
\noindent 
Recently, operator quantum error-correcting codes have been proposed
to unify and generalize decoherence free subspaces, noiseless
subsystems, and quantum error-correcting codes. This note introduces a
natural construction of such codes in terms of Clifford codes, an
elegant generalization of stabilizer codes due to
Knill. Character-theoretic methods are used to derive a simple method
to construct operator quantum error-correcting codes from any
classical additive code over a finite field, which obviates the need for
self-orthogonal codes. 
\end{abstract}

\paragraph{Introduction.} 
One of the main challenges in quantum information processing is the
protection of the quantum information against various sources of
errors. A possible remedy is given by encoding the quantum information
in a subspace $C$ of the state space $H$ of the quantum system.  If
such a quantum error-correcting code $C$ is well-chosen, then many
errors can be corrected through active recovery operations.  A more
recent development is the encoding of quantum information into a
subsystem $A$ of the state space \cite{kribs05,kribs05b}. This means
that $C$ is further decomposed into a tensor product of vector spaces
$A$ and $B$ such that
$$ H = C\oplus C^\perp = (A\otimes B) \oplus C^\perp.$$ One refers
to $C$ as an operator quantum error-correcting code with subsystem $A$
and co-subsystem $B$. Some authors refer to the co-subsystem as the
gauge subsystem. One advantage is that errors affecting the
co-subsystem $B$ alone do not require any active
error-correction. Furthermore, one can detect all errors that map the
encoded information into the orthogonal complement $C^\perp$ of $C$.

The operator quantum error-correcting codes generalize and unify the
main methods of passive and active quantum error-correction:
decoherence free subspaces, noiseless subsystems, and quantum
error-correcting codes. More background on operator quantum
error-correcting codes can be found, for example, in
references~\cite{bacon06,knill06,kribs05,kribs05b,kribs05c,poulin05}.

The purpose of this paper is to introduce a natural method for
constructing such operator quantum error-correcting codes.  Our
approach is based on an elegant formalism to construct quantum
error-correcting codes that has been introduced in 1996 by Knill
as a generalization of the stabilizer code concept.  At the heart of
this quantum code construction is a famous theorem by Clifford
concerning the restriction of irreducible representations of finite
groups to normal subgroups, so we referred to these codes as
``Clifford codes'' in \cite{klappenecker031,klappenecker033}, although
``Knill codes'' is perhaps a more appropriate name.  Unexpectedly, it
turned out that Clifford codes are in many cases stabilizer codes, so
this construction did not become as widely known as it should.

In our approach, we construct a Clifford code $C$ and give conditions
that ensure that this code decomposes into a tensor product $C=
A\otimes B$. The Clifford codes allow us to control the dimensions of
$A$ and $B$, and we get a simple characterization of the detectable
errors of the operator quantum error-correcting code.  Since there may
exist many different ways to construct the same Clifford code $C$, we
should note that these constructions can lead to different tensor
product decompositions. In fact, even if one is just interested in the
tensor decomposition of a stabilizer code $C$, then the Clifford codes
can provide a natural way to induce an operator quantum
error-correcting code on $C$.
\smallskip

{\small \textit{Notation.} If $N$ is a group, then $Z(N)$ denotes the
center of $N$.  We denote by $\Irr(N)$ the set of irreducible
characters of $N$. If $\chi$ and $\psi$ are characters of $N$, then
$(\chi,\psi)_N=|N|^{-1}\sum_{n\in N} \chi(n)\psi(n^{-1})$ defines a
scalar product on the vector space of class functions on $N$, and
$\Irr(N)$ is an orthonormal basis of this space. We denote by $\supp(\chi)=\{ n\in N|\, \chi(n)\neq 0\}$.
If $\chi\in \Irr(N)$,
then $Z(\chi)=\{ n\in N\,|\, \chi(1)=|\chi(n)|\}$ denotes the
quasikernel of $\chi$. Suppose that $G$ is a group that contains $N$
as a subgroup.  If $\phi\in \Irr(G)$, then $\phi_N$ denotes the
restriction of this character to $N$. If $x,y\in N$, then
$[x,y]=x^{-1}y^{-1}xy$ is the commutator. If $A$ and $B$ are subgroups
of a group, then $[A,B]=\langle [a,b]\,|\, a,\in A \text{ and } b\in
B\rangle$ is the commutator subgroup of $A$ and $B$. In particular,
$N'=[N,N]$ denotes the derived subgroup of $N$.  The reader can find
background material on finite groups in \cite{robinson95} and on
character theory in \cite{isaacs94}.}

\paragraph{Clifford Codes.} 
Before introducing the concept of a Clifford code, we need to fix a
notion of errors that generalizes the concept of the Pauli group. We
say that a finite group $E$ is an abstract error group if it has a
faithful irreducible unitary representation $\rho$ of degree
$d = |E:Z(E)|^{1/2}$. The irreducibility of the representation ensures that
one can express any error acting on $\C^d$ as a linear
combination of the matrices $\rho(g)$, with $g\in E$. The fact that
the representation is faithful and has the largest possible degree
ensures that the set of matrices $\{ \rho(g)\,|\, g\in T\}$, where $T$
is a set of representatives of $E/Z(E)$, forms a \textit{basis}\/ of
the vector space of $d\times d$ matrices.

A Clifford code is constructed with the help of a normal subgroup $N$
of the error group $E$ and an irreducible character $\chi$ of $N$.
Let $\phi$ denote the irreducible character corresponding to the
representation $\rho$ of the group $E$, that is, $\phi(g)=\Tr \rho(g)$
for $g\in E$. Suppose that $N$ is a normal subgroup of $E$ and that
$\chi$ is an irreducible character of $N$ such that $(\chi,\phi_N)_N>
0$.  Then the Clifford code $C$ corresponding to $(E,\rho,N,\chi)$ is
defined as the image of the orthogonal projector
$$ P = \frac{\chi(1)}{|N|} \sum_{n\in N} \chi(n^{-1})\rho(n),$$
see~\cite[Theorem~1]{klappenecker033}. We emphasize that if we refer
to a Clifford code with data $(E,\rho,N,\chi)$, then it is assumed that
$(\chi,\phi_N)>0$, as this condition ensures that $\dim C>0$.

Recall that an error $e$ in $E$ is detectable by the quantum code $C$
if and only if $P\rho(e)P=\lambda_e P$ holds for some $\lambda_e\in \C$.

The image of $P$ is the homogeneous component that consists of the
direct sum of all irreducible $\C N$-submodules with character $\chi$
that are contained in the restriction of $\rho$ to~$N$.  The elements
$e$ in $E$ that satisfy $\rho(e)C= C$ form a group known as
the inertia group $I_E(\chi)= \{ g\in E\,|\, \chi(gxg^{-1})=\chi(x)
\text{ for all } x\in N\}.$ We note that $C$ is an irreducible
$\C[I_E(\chi)]$-module. Let $\vartheta$ be the irreducible character
corresponding to this module.

\begin{fact}
\label{fact:clifford}
Let $C$ be a Clifford code with data $(E,\rho,N,\chi)$. Then the
dimension of the code is given by $\dim C = |Z(E)\cap N| |E:
Z(E)|^{1/2}\chi(1)^2/|N|$.  An error $e$ in $E$ can be detected by $C$
if and only if $e$ is in $E-(I_E(\chi)-Z(\vartheta))$.
\end{fact}

For a proof of this fact see \cite{klappenecker033} and for more
background on Clifford codes see~\cite{klappenecker031} and the
seminal papers~\cite{knill96a,knill96b}.

\paragraph{Operator Quantum Error-Correcting Codes.} 
We are now concerned with the construction of a  decomposition of the
Hilbert space $H$ in the form
$$ H = (A\otimes B) \oplus C^\perp.$$ Put differently, we seek a
decomposition of the Clifford code $C$ as a tensor product $A\otimes
B$.

The next theorem gives a construction of operator quantum
error-cor\-recting codes when one can express the inertia group
$I_E(\chi)$ as a central product $I_E(\chi)=LN$, where $L$ is a
subgroup of $E$ such that $[L,N]=1$.

\begin{theorem}\label{th:first}
Suppose that $C$ is a Clifford code with data $(E,\rho, N,\chi)$.  If
the inertia group $I_E(\chi)$ is of the form $I_E(\chi)=LN$, where $L$
is a subgroup of $E$ such that $[L,N]=1$, then $C$ is an operator
quantum error-correcting code $C= A\otimes B$ such that
\begin{compactenum}[i)]
\item $\dim A = |Z(E)\cap N| |E: Z(E)|^{1/2}\chi(1)/|N|$,
\item $\dim B=\chi(1)$.
\end{compactenum}
The subsystem $A$ is an irreducible $\C L$-module 
with character $\chi_A\in \Irr(L)$. 
An error $e$ in $E$ is detectable by
subsystem $A$ if and only if $e$ is contained in the set
$E-(I_E(\chi)-Z(\chi_A)N)$.
\end{theorem}
\begin{proof}
Since the Clifford code $C$ is an irreducible $\C[I_E(\chi)]$-module
and $I_E(\chi)=LN$, with $[L,N]=1$, there exists an irreducible $\C
L$-module $A$ and an irreducible $\C N$-module $B$ such that $C\cong
A\otimes B$, see~\cite[Proposition 9.14]{GLS2}. If $\chi_A\in \Irr(L)$
is the character associated with the module $A$, $\chi_B\in \Irr(N)$
the character associated with $B$, and $\vartheta\in \Irr(I_E(\chi))$
the character associated with $C$, then $\vartheta$ is of the form
$\vartheta(\ell n)=\chi_A(\ell)\chi_B(n)$ with $\ell \in L$ and $n\in
N$.

As the restriction of $C$ to a $\C N$-module contains an irreducible
$\C N$-module $W$ with character $\chi$, we must have
$$ \begin{array}{lcl}
\displaystyle 
(\vartheta_N , \chi)_N = \frac{1}{|N|} \sum_{n\in N}
\vartheta(1,n^{-1}) \chi(n) &=& 
\displaystyle \frac{1}{|N|} \sum_{n\in N}
\chi_A(1)\chi_B(n^{-1}) \chi(n) \\ 
&=& \chi_A(1)  (\chi_B, \chi)_N > 0.
   \end{array}
$$ 
Since $\Irr(N)$ forms an orthonormal basis with
respect to $(\,\cdot\,,\,\cdot\,)_N$, we
can conclude that the irreducible character $\chi_B$ must be equal to
$\chi$. It follows that $C\cong A\otimes W$.

The dimension of $W\cong B$ is $\chi(1)$, and by Fact~\ref{fact:clifford} 
the dimension of $C$ is given by 
$$ \Tr P = |Z(E)\cap N| |E:Z(E)|^{1/2} \chi(1)^2/|N|.$$
The dimension of $B$ follows from the formula $\dim C=\dim A\dim B$. 

An error $e\in E-I_E(\chi)$ maps $C$ to an orthogonal complement, so the
errors are detectable. An error $e$ in $Z(\chi_A)N$ acts by
scalar multiplication on $A$ and arbitrarily on $B$, so these errors
are by definition detectable (harmless would be a better
word). Therefore, all errors in $E-(I_E(\chi)-Z(\chi_A)N)$ are
detectable. Conversely, an error $e$ in $I_E(\chi)-Z(\chi_A)N$
cannot be detectable, since $e$ does not act by scalar multiplication
on $A$, and thus does not preserve the encoded quantum information.
\end{proof}

The data given in the previous theorem can be easily computed,
especially with the help of a computer algebra system such as GAP or
MAGMA. 

We will now consider some important special cases.  Recall that most
abstract error groups that are used in the literature satisfy the
constraint $E'\subseteq Z(E)$ (put differently, the quotient group
$E/Z(E)$ is abelian). In that case, we are able to obtain a
characterization of the resulting operator quantum error-correcting
codes that does not depend on the choice of the character $\chi$.

\begin{theorem}\label{th:second} 
Suppose that $E$ is an abstract error group such that $E'\subseteq Z(E)$. 
Suppose that $C$ is a Clifford code with data $(E,\rho, N,\chi)$.  
In this case, the inertia group is given by $I_E(\chi)=C_E(Z(N))$. 
If $C_E(Z(N))=LN$ for some subgroup $L$ of $E$ such that $[L,N]=1$, 
then $C$ is an operator
quantum error-correcting code $C= A\otimes B$ such that
\begin{compactenum}[i)]
\item $\dim A = |Z(E)\cap N| |E: Z(E)|^{1/2}|N:Z(N)|^{1/2}/|N|$,
\item $\dim B=|N:Z(N)|^{1/2}$.
\end{compactenum}
An error $e$ in $E$ is detectable by
subsystem $A$ if and only if $e$ is contained in the set
$E-(C_E(Z(N))-Z(L)N)$.
\end{theorem}
\begin{proof}
Since the abstract error group $E$ satisfies the condition
$E'\subseteq Z(E)$, the inertia group of the character $\chi$ in $E$
can be fully determined; it is given by $T:=I_E(\chi)=C_E(Z(N))$, see
\cite[Lemma~5]{klappenecker033}. 

Suppose that 
$$ P_1 = \frac{\chi(1)}{|N|}\sum_{n\in N} \chi(n^{-1})\rho(n)$$ is the
orthogonal projector onto $C$. The assumption $E'\subseteq Z(E)$
implies that there exists a linear character $\varphi$ of $\Irr(Z(N))$
such that 
$$ P_2 = \frac{1}{|Z(N)|}\sum_{n\in Z(N)} \varphi(n^{-1})\rho(n)$$
satisfies $P_1=P_2$, see \cite[Theorem~6]{klappenecker033}.

Let $\phi$ be the character of the representation $\rho$, that is,
$\phi(g)=\Tr \rho(g)$ for $g\in E$. We have $ \Tr P_1 = \chi(1)^2
\phi(1)|N\cap Z(E)|/|N|$ and $\Tr P_2 = \phi(1)|N\cap
Z(E)|/|Z(N)|$. Since $P_1=P_2$ project onto the code space $C$, and
$\dim C>0$, we have $\Tr P_1/\Tr P_2=1$, which implies
$\chi(1)^2=|N\colon Z(N)|$. Therefore, the claims i) and ii) follow
from Theorem~\ref{th:first}.

Let $\vartheta\in \Irr(T)$ be the character associated with the 
$\C[T]$-module $C$; put differently, $\vartheta$ is the unique
character in $\Irr(T)$ that satisfies $(\vartheta_N,\chi)_N>0$ and
$(\phi_T,\vartheta)_T>0$. Since $Z(E)\le T$ and
$(\phi_T,\vartheta)_T>0$, it follows from Lemma~\ref{l:support2} that
$\supp(\vartheta)=Z(T)$. 

Since the inertia group $T$ is a central product given by $T=LN$ with
$[L,N]=1$, there exist characters $\chi_A\in \Irr(L)$ and
$\chi_B=\chi\in \Irr(N)$ such that $\vartheta(\ell
n)=\chi_A(\ell)\chi(n)$ for $\ell\in L$ and $n\in N$.  By
Lemma~\ref{l:center}, we have $Z(T)=Z(L)Z(N)$; thus,
$\supp(\vartheta)=Z(L)Z(N)$. This implies that $\supp(\chi_A)=L\cap
Z(L)Z(N)=Z(L)$; hence $Z(\chi_A)=Z(L)$.  The characterization of the
detectable errors is obtained by substituting these facts in
Theorem~\ref{th:first}.
\end{proof}

In the previous theorem, we still need to check whether $C_E(Z(N))$
decomposes into a central product of $N$ and some group $L$. In the
case of extraspecial $p$-groups (which is arguably the most popular
choice of abstract error groups) the decomposition of the inertia
group into a central product is always guaranteed, as we will show
next.

Recall that a finite group $E$ whose order is a power of a prime $p$
is called extraspecial if its derived subgroup $E'$ and its center
$Z(E)$ coincide and have order $p$. An extraspecial $p$-group is an
abstract error group. The quotient group $\overline{E}=E/Z(E)$ is the
direct product of two isomorphic elementary abelian
$p$-groups. Therefore, one can regard $\overline{E}$ as a vector space
$\F_p^{2n}$ over the finite field $\F_p$.

Let $\zeta$ be a fixed generator of the cyclic group $Z(E)$.  As the 
commutator $[x,y]$ depends only on the cosets $\overline{x}=xZ(E)$ and
$\overline{y}=yZ(E)$, one can determine a well-defined function $s\colon
\overline{E}\times \overline{E}\rightarrow \F_p$ by
$[x,y]=\zeta^{s(\overline{x},\overline{y})}$. The function $s$ is a
nondegenerate symplectic form. We note that two elements $x$ and $y$
in $E$ commute if and only if $s(\overline{x},\overline{y})=0$.
We write $\overline{x}\, \sdual\, \overline{y}$ if and only if 
$s(\overline{x},\overline{y})=0$.

For a subgroup $G$ of $E$, we will use $\overline{G}$ to denote $G/Z(E)$. 
\begin{lemma}\label{l:inertia}
If $E$ is an extraspecial $p$-group and $N$ a normal subgroup of $E$,
then $C_E(Z(N))=NC_E(N)$.
\end{lemma}
\begin{proof}
Since $Z(E)\le NC_E(N)\le C_E(Z(N))$, it suffices to show that the
dimensions of the $\F_p$-linear vector spaces 
$$\overline{NC_E(N)} \quad \text{and}\quad \overline{C_E(Z(N))}$$ are
the same. Suppose that $z=\dim\overline{Z(N)}$ and
$k=\dim\overline{N}$. 
Then
$$
\begin{array}{lcl}
\dim \ol{NC_E(N)}&=& \dim (\ol{N}+\ol{N}^\sdual)
= \dim\ol{N}+\dim\ol{N}^\sdual-\dim(\ol{N}\cap \ol{N}^\sdual)\\
&=& \dim\ol{N}+\dim\ol{N}^\sdual-\dim(\ol{Z(N)})\\
&=& k+ (2n-k)-z=2n-z,
  \end{array}
$$ which coincides with $\dim
\ol{C_E(Z(N))}=\dim\ol{Z(N)}^\sdual=2n-z$, and this proves our claim.
\end{proof}

The next theorem shows that it suffices to choose a normal subgroup
$N$ of the extraspecial $p$-group $E$, and this choice determines the
parameters of an operator quantum error-correcting code provided by a
Clifford code $C$.

\begin{theorem}\label{th:third} 
Suppose that $E$ is an extraspecial $p$-group.  If $C$ is a Clifford 
code with data $(E,\rho, N,\chi)$, with $N\neq 1$, 
then $C$ is an operator quantum
error-correcting code $C= A\otimes B$ such that
\begin{compactenum}[i)]
\item $\dim A = |Z(E)\cap N| |E: Z(E)|^{1/2}|N:Z(N)|^{1/2}/|N|$,
\item $\dim B=|N:Z(N)|^{1/2}$.
\end{compactenum}
An error $e$ in $E$ is detectable by
subsystem $A$ if and only if $e$ is contained in the set
$E-(NC_E(N)-N)$.
\end{theorem}
\begin{proof}
The inertia group $I_\chi(E)=C_E(Z(N))$, since $E'\subseteq Z(E)$,
see~\cite[Lemma~5]{klappenecker033}. By Lemma~\ref{l:inertia}, we have
$I_E(\chi)=LN=NL$ with $L=C_E(N)$. Thus, $C$ is an operator quantum
error-correcting code and the statements i) and ii) follow from
Theorem~\ref{th:second}. Furthermore, Theorem~\ref{th:second} shows
that an error $e$ in $E$ is detectable if and only if $e\in
E-(NC_E(N)-Z(L)N)$. Since $E$ is a $p$-group and $N\neq 1$, we have
$N\cap Z(E)\neq 1$; hence $Z(E)\le N$. We note that
$\ol{Z(L)}\subseteq \ol{L}\cap \ol{L}^\sdual=
\ol{N}^\sdual\cap\ol{N}\subseteq \ol{N}$; therefore, $N\subseteq Z(L)N\subseteq Z(N)N=N$, forcing $Z(L)N=N$.
\end{proof}

\paragraph{Classical Codes.} 
We conclude this note by showing how the previous results can be
related to classical coding theory.

Let $a$ and $b$ be elements of the finite field $\F_q$ of characteristic $p$.  
We define unitary operators $X(a)$ and $Z(b)$ on~$\C^q$ by
$$ X(a)\ket{x}=\ket{x+a},\qquad Z(b)\ket{x}=\omega^{\tr(bx)}\ket{x},$$
where $\tr$ denotes the trace operation from the extension field
$\F_q$ to the prime field $\F_p$, and $\omega=\exp(2\pi i/p)$ is a
primitive $p$th root of unity.  Let $\mathbf{a}=(a_1,\dots, a_n)\in
\F_q^n$. We write $ X(\mathbf{a}) = X(a_1)\otimes\, \cdots \,\otimes
X(a_n)$ and $Z(\mathbf{a}) = Z(a_1)\otimes\, \cdots \,\otimes Z(a_n)$
for the tensor products of $n$ error operators. One readily checks that 
the group
$$ E = \langle X(a), Z(b)\,|\, a, b\in \F_q^n\rangle$$ is an
extraspecial $p$-group of order $pq^{2n}$. As a representation $\rho$,
we can take the identity map on $E$. We have $E/Z(E)\cong \F_q^{2n}$. 

We need to introduce a notion of weights of errors.  Recall that an
error in $E$ can be expressed in the form $\alpha X(a)Z(b)$ for some
nonzero scalar $\alpha$. The weight of $\alpha X(a)Z(b)$ is defined as
$|\{ i\,|\, 1\le i\le n, a_i\neq 0 \text{ or } b_i\neq 0\}|$, that is,
as the number of quantum systems that are affected by the
error. Similarly, we can introduce a weight on vectors of $\F_q^{2n}$
by
$$ \swt(a|b)=\{ i\,|\, 1\le i\le n, a_i\neq 0 \text{ or } b_i\neq 0\}|$$
for $a,b\in \F_q^n$.

Theorem~\ref{th:third} suggests 
the following approach to construct 
operator quantum error-correcting codes. 
\begin{theorem}\label{th:four}
Let $X$ be a classical additive subcode of\/ $\F_q^{2n}$ such that
$X\neq \{0\}$ and let $Y$ denote its subcode $Y=X\cap X^\sdual$. 
If $x=|X|$ and $y=|Y|$, then
there exists an operator quantum error-correcting code $C=
A\otimes B$ such that
\begin{compactenum}[i)]
\item $\dim A = q^n/(xy)^{1/2}$, 
\item $\dim B = (x/y)^{1/2}$.
\end{compactenum}
The minimum distance of subsystem $A$ is given by
$d=\swt((X+X^\sdual)-X)=\swt(Y^\sdual-X)$. Thus, the subsystem $A$ can
detect all errors in $E$ of weight less than $d$, and can correct all
errors in $E$ of weight $\le \lfloor (d-1)/2\rfloor$.
\end{theorem}
\begin{proof}
Let $E$ be the extraspecial $p$-group of order $pq^{2n}$, and let $N$
be the full preimage of $\ol{N}=X$ in $E$ under the canonical quotient
map.  Therefore, we can apply Theorem~\ref{th:third}. The remainder of
the proof justifies how the parameters given in Theorem~\ref{th:third}
can be expressed in terms of the code sizes $x$ and $y$.

Then $\ol{Z(N)}=X\cap X^\sdual=Y$.  By definition, $N$
contains $Z(E)$; hence, $Z(E)\le Z(N)$. 
It follows that
$|N:Z(N)|=|\overline{N}:\ol{Z(N)}|=x/y$, so ii) follows from
Theorem~\ref{th:third}. For the claim i), we remark that
$x=|X|=|N|/p$, which implies that $\dim A =(p/|N|)
|E:Z(E)|^{1/2}|N:Z(N)|^{1/2}=q^n(x/y)^{1/2}/x$.

The minimum distance of subsystem $A$ is the weight of the smallest
nondetectable error, so it is the minimum weight of an error in the
set $NC_E(N)-N=C_E(Z(N))-N$. Since the quotient map $E\rightarrow
\ol{E}$ maps an error $e$ of weight $w$ onto a vector $\ol{e}$ such
that $w=\swt{\ol{e}}$, the claim about the minimum distance follows
from the observations that $\ol{NC_E(N)-N}=(X+X^\sdual)-X$ and 
$\ol{C_E(Z(N))-N}=Y^\sdual-X$. 
\end{proof}

An operator quantum error-correcting code with parameters
$((n,a,b,d))_q$ is a subspace $C=A\otimes B$ of a $q^n$-dimensional
Hilbert space $H$ such that $a=\dim A$, $b=\dim B$, and the subsystem
$A$ has minimum distance $d$. The above theorem constructs an
$((n,q^n/(xy)^{1/2},(x/y)^{1/2},d))_q$ operator quantum
error-correcting code given a classical $(n,x)_q$ code $X$ and its
$(n,y)_q$ subcode $Y=X\cap X^{\sdual}$.
We write $[[n,k,r,d]]_q$ for an
$((n,q^k,q^r,d))_q$ operator quantum error-correcting code.

Sometimes one would like to characterize the minimum distance in terms
of the familiar Hamming weight. For this purpose, we reformulate
the above result in terms of codes of length $n$ over $\F_{q^2}$.

Let $(\beta,\beta^q)$ be a fixed normal basis of $\F_{q^2}$ over
$\F_q$. We can define a bijection $\phi$ from $\F_q^{2n}$ onto
$\F_{q^2}^n$ by setting
$$\phi((a|b))=\beta a+\beta^q b\quad \text{for}\quad (a|b)\in
\F_q^{2n}.$$ The map is chosen such that a vector $(a|b)$ of
symplectic weight $x$ is mapped to a vector $\phi((a|b))$ of Hamming
weight $x$. If we define a trace-alternating form 
$\langle v|w\rangle_a$ for vectors $v$ and $w$ in $\F_{q^2}^n$ by 
$$\langle v|w\rangle_a=\tr_{q/p}\left(\frac{v\cdot w^q-v^q\cdot
w}{\beta^{2q}-\beta^q}\right),$$ then it is easy to show that $
\langle c|d\rangle_s = \langle \phi(c)|\phi(d)\rangle_a$ holds for all
$c,d\in \F_q^{2n}$, see~\cite[Lemma 14]{pre3}.  Specifically, we have
$c\, \sdual\, d$ if and only if $\phi(c)\, \adual\, \phi(d)$.
Therefore, the previous theorem can be reformulated terms of codes of
length $n$ over $\F_{q^2}$ as follows:

\begin{theorem}\label{th:five}
Let $X$ be a classical additive subcode of\/ $\F_{q^2}^{n}$ such that
$X\neq \{0\}$ and let $Y$ denote its subcode $Y=X\cap X^\adual$. 
If $x=|X|$ and $y=|Y|$, then
there exists an operator quantum error-correcting code $C=
A\otimes B$ such that
\begin{compactenum}[i)]
\item $\dim A = q^n/(xy)^{1/2}$, 
\item $\dim B = (x/y)^{1/2}$.
\end{compactenum}
The minimum distance of subsystem $A$ is given by
$$d=\wt((X+X^\adual)-X)=\wt(Y^\adual-X),$$
where $\wt$ denotes the Hamming weight.  Thus, the subsystem $A$ can
detect all errors in $E$ of Hamming weight less than $d$, and can correct all
errors in $E$ of Hamming weight $\lfloor (d-1)/2\rfloor$ or less. 
\end{theorem}
\begin{proof}
This follows from Theorem~\ref{th:four} and the definition of the
isometry~$\phi$.
\end{proof}

The above connections of Clifford operator quantum error-correcting
codes to classical code allow one to explore a plethora of code
constructions.

\paragraph{Conclusions.} 
We have introduced a method for constructing operator quantum
error-correcting codes. We have seen that a Clifford codes $C$ offers
naturally a tensor-product decomposition $C= A\otimes B$, where
the dimensions of the subsystems are controlled by the choice of the
normal subgroup $N$ and its character $\chi$. 

Our construction in terms of classical codes is fairly simple: Any
classical (additive) code over a finite field can be used to construct
an operator quantum error-correcting code. In particular, we do not
require any self-orthogonality conditions as in the case of stabilizer
code constructions. 

The most prominent open problem concerning operator quantum
error-correcting codes is whether one can achieve better error
correction that by means of a quantum error-correcting code. The
construction given in Theorem~\ref{th:four} allows one to compare the
parameters of Clifford codes with the parameters of stabilizer codes.
One should note that a fair comparison should be made between
$[[n-r,k,d]]$ stabilizer codes and $[[n,k,r,d]]$ Clifford codes.  It
would be helpful to have bounds on the best possible minimum distance
$d$ of Clifford codes to answer this question. 

\paragraph{Acknowledgments.}
This research was supported by NSF grant CCF-0218582, NSF CAREER award
CCF-0347310, and a TITF project.

\appendix
\section{Appendix}  
In this appendix, we prove some simple technical results on groups and
characters.

\begin{lemma}\label{l:support}
Let $E$ be a finite group such that $E'\subseteq Z(E)$, and let $H$ be
a subgroup of $E$. If $\chi\in \Irr(H)$ satisfies $Z(E)\cap \ker \chi
= \{1\}$, then $\supp \chi = Z(H)$.
\end{lemma}
\begin{proof}
Let $h\in \supp(\chi)$. Seeking a contradiction, we assume that $h\in
H-Z(H)$. Since $E'\subseteq Z(E)$, there exists an element $g\in H$
such that $ghg^{-1}=zh$ with $z\in Z(E)$ such that $z\neq 1$.  Since 
$zh\in H$ and $h\in H$, we have $z\in H\cap Z(E)$. As $\chi$ is
irreducible, the element $z\in H\cap Z(E)$ is represented by $\omega
I$ for some $\omega\in \C$ by Schur's lemma; furthermore, $\omega \neq
1$, since $Z(E)\cap \ker \chi=\{1\}$.
We note that 
$ \chi(h)=\chi(ghg^{-1}) = \chi(zh)= \omega\chi(h)$, with
$\omega\neq 1$, forcing $\chi(h)=0$, contradiction. 

The elements of $Z(H)$ belong to the support of $\chi$, since they are
represented by scalar invertible matrices.
\end{proof}

\begin{lemma}\label{l:support2}
Let $E$ be a finite group such that $E'\subseteq Z(E)$, and let
$\phi\in \Irr(E)$ be a faithful character of degree
$\phi(1)=|E:Z(E)|^{1/2}$. Let $T$ be a subgroup of $E$ such that $Z(E)\le T$. 
If $\vartheta\in \Irr(T)$ and $(\phi_T,\vartheta)_T>0$, then 
$\supp(\vartheta)=Z(T)$. 
\end{lemma}
\begin{proof}
Let $Z=Z(E)$. We have $\supp(\phi)=Z$ by \cite[Lemma 2.29]{isaacs94}. 
Since the support of $\phi$ equals $Z$, it follows from the definitions that 
$$ 0<(\phi_T,\vartheta)_T = \frac{1}{|T:Z|}(\phi_Z,\vartheta_Z)_Z.$$
Clearly, $\phi_Z=\phi(1)\varphi$ and $\vartheta_Z=\vartheta(1)\theta$ 
for some linear characters $\varphi$
and 
$\theta$ of $Z$. As
$(\phi_Z,\vartheta_Z)_Z=\phi(1)\vartheta(1)(\varphi,\theta)_Z>0$, we
must have $\theta=\varphi$. Since $\phi$ is faithful, it follows that
$\varphi=\theta$ is faithful; hence, $\ker \vartheta \cap Z(E)=\{1\}$.
Thus, $\supp\vartheta=Z(T)$ by Lemma~\ref{l:support}.
\end{proof}

\begin{lemma}\label{l:center} 
Suppose that $T$ is a group with subgroups $L$ and $N$ such that
$T=LN$ and $[L,N]=1$. Then $Z(T)=Z(L)Z(N)$. 
\end{lemma}
\begin{proof} 
Since $T=LN$, an arbitrary element $z$ of $Z(T)$ can be expressed in
the form $z=ln$ for some $l\in L$ and $n\in N$.  For $n'$ in $N$, we
have $lnn'=n'ln=ln'n$, where the latter equality follows from
$[L,N]=1$. Consequently, $nn'=n'n$ for all $n'$ in $N$, so $n$ is an
element of $Z(N)$. Similarly, $l$ must be an element of $Z(L)$.  It
follows that $Z(T)=Z(L)Z(N)$.
\end{proof}


\begin{thebibliography}{10}
\bibitem{bacon06}
D.~Bacon.
\newblock Operator quantum error correcting subsystems for self-correcting
  quantum memories.
\newblock {\em Phys. Rev.~A}, 73(012340), 2006.

\bibitem{GLS2}
D.~Gorenstein, R.~Lyons, and R.~Solomon.
\newblock {\em The Classification of the Finite Simple Groups, Number 2},
  volume~40 of {\em Mathematical Surveys and Monographs}.
\newblock AMS, 1994.

\bibitem{isaacs94}
I.M. Isaacs.
\newblock {\em Character Theory of Finite Groups}.
\newblock Dover, 1994.

\bibitem{pre3}
A.~Ketkar, A.~Klappenecker, S.~Kumar, and P.K. Sarvepalli.
\newblock Nonbinary stabilizer codes over finite fields.
\newblock To appear in IEEE Trans. Inform. Theory, November, 2006.

\bibitem{klappenecker033}
A.~Klappenecker and M.~R{\"otteler}.
\newblock Beyond stabilizer codes {II}: {C}lifford codes.
\newblock {\em IEEE Transaction on Information Theory}, 48(8):2396--2399, 2002.

\bibitem{klappenecker031}
A.~Klappenecker and M.~R{\"o}tteler.
\newblock Clifford codes.
\newblock In R.~Brylinski and G.~Chen, editors, {\em Mathematics of Quantum
  Computing}, pages 253--273. Chapman \& Hall/CRC Press, 2002.

\bibitem{knill96b}
E.~Knill.
\newblock Group representations, error bases and quantum codes.
\newblock Los Alamos National Laboratory Report LAUR-96-2807, 1996.

\bibitem{knill96a}
E.~Knill.
\newblock Non-binary unitary error bases and quantum codes.
\newblock Los Alamos National Laboratory Report LAUR-96-2717, 1996.

\bibitem{knill06}
E.~Knill.
\newblock On protected realizations of quantum information.
\newblock Eprint: quant-ph/0603252, 2006.

\bibitem{kribs05c}
D.~W. Kribs.
\newblock A brief introduction to operator quantum error correction.
\newblock Eprint: math/0506491, 2005.

\bibitem{kribs05}
D.~W. Kribs, R.~Laflamme, and D.~Poulin.
\newblock Unified and generalized approach to quantum error correction.
\newblock {\em Phys. Rev. Lett.}, 94(180501), 2005.

\bibitem{kribs05b}
D.~W. Kribs, R.~Laflamme, D.~Poulin, and M.~Lesosky.
\newblock Operator quantum error correction.
\newblock Eprint: quant-ph/0504189, 2005.

\bibitem{poulin05}
D.~Poulin.
\newblock Stabilizer formalism for operator quantum error correction.
\newblock {\em Phys. Rev. Lett.}, 95(230504), 2005.

\bibitem{robinson95}
D.J.S. Robinson.
\newblock {\em A Course in the Theory of Groups}.
\newblock Springer, 2nd edition, 1995.

\end{thebibliography}

\end{document}